\documentclass[a4paper]{article}
\usepackage[T1]{fontenc}
\usepackage[utf8x]{inputenc}
\usepackage[english]{babel}
\usepackage{lmodern} 
\usepackage[colorlinks=true, allcolors=blue]{hyperref}
\newcommand{\email}[1]{\href{mailto:#1}{\tt{\nolinkurl{#1}}}}
\newcommand{\orcid}[1]{ORCID: \href{https://orcid.org/#1}{\tt{\nolinkurl{#1}}}}
\usepackage[sfdefault,lf]{carlito}
\usepackage[parfill]{parskip}

\usepackage{fancyhdr}
\usepackage{natbib}
\usepackage{authblk}
\usepackage[a4paper,top=3cm,bottom=2cm,left=3cm,right=3cm,marginparwidth=1.75cm]{geometry}

\usepackage{amsmath}
\usepackage{graphicx}
\usepackage{booktabs}
\usepackage{comment}
\usepackage[colorinlistoftodos]{todonotes}
\graphicspath{ {./figures/} }
\def\W{Wider\o e}

\def\BT{Bruno Touschek}
\def\RW{Rolf Wider\o e}
\def\W{Wider\o e}
\def\PM{Pierre Marin}
\def\LAL{Laboratoire de l'Acc\'el\'erateur Lin\'eaire}

\def\WW2{Second World War II}

\def\Gott{G\"ottingen}

\def\Gott{G\"ottingen}

\def\tr{\textcolor{red}}

\def\bef{\begin{figure}}
\def\enf{\end{figure}}
\def\befoot{\begin{footnotesize}}
\def\enfoot{\end{footnotesize}}

\def\RSUPD{Rome Sapienza University, Physics Department Archives}

\def\JH{Jacques Ha\"issinski}
\def\PM{Pierre Marin}

\def\tr{\textcolor{red}}

\title{Particle accelerators as   a path to  intergovernmental 
collaboration: the case of  the first electron positron collider and the power  of  theoretical physics}
\author{ Giulia Pancheri}

\affil{INFN Frascati National Laboratories, Via Enrico Fermi 56, Frascati, I00044, Italy, giulia.pancheri@lnf.infn.it}

\date{\today}
  
\begin{document}
\maketitle
\thispagestyle{fancy}
\begin{abstract} I revisit the history of the first matter-antimatter collider built in Italy in 1960 and  carried to success through an inter-European collaboration with France, 
{to underline the need for large intergovernmental efforts in the realization of  particle accelerators such as the proposed electron synchrotron 
in the Greater  Caribbean region. }  The foundational role of the Austrian born theoretical physicist Bruno Touschek and his  
 life path through Europe before and after World War II  is briefly recalled.  The role of CERN {as a post-war intergovernmental institution} in jump-starting the international collaboration between the Frascati National Laboratories in Italy and the \LAL \ d'Orsay in France is highlighted through historical documents, and contributions from interviews with the  protagonists of Touschek's  game-changing proposal.\end{abstract}

keywords: history of physics, elementary particles, electron-positron colliders
\maketitle

Preprint No.: INFN – 2026-06-03-LNF\\

\section{Introduction}
\label{sec:intro}

{The effort to promote the construction of an electron synchrotron for the Greater Caribbean region \citep{Castano:2024,castano2026cities} highlights the role of particle accelerator facilities not only as fundamental tools for scientific research, but also as infrastructures capable of contributing to political integration and to advances in society along multiple dimensions.\footnote{ {A recent opinion comment highlights how 
investing in a synchrotron light source means committing to scientific sovereignty, competitiveness, and the region’s future,
 \url{https://acento.com.do/opinion/soberania-cientifica-en-el-gran-caribe-cinco-anos-de-gclsi-9696387.html}.}}
 This initiative must also be viewed in the broader historical context of Latin America’s complex, but ultimately successful, engagement with Big Science. The introduction of large-scale scientific research into the Andean  region was neither straightforward nor free of obstacles, yet it produced significant experiences whose impact extended well beyond the scientific community. An early example was the emergence of high-energy physics in the Andean region discussed in a seminal article of more than forty years ago \citep{Cini:1983}, which contributed to stimulating a debate on the role of advanced science in the region and helped inspire a broader policy approach that would subsequently be led by the Centro Internacional de Física (CIF), 
  making Latin American scientists regular participants in major international facilities.}\footnote{{ See  December 1993 
{\it  Report of the Steering Committee for Fundamental Physics},
 \url{http://researchgate.net/publication/374730155\_Report\_of\_the\_Steering\_Committee\_for\_Fundamental\_Physics}, 
 DOI:10.13140/RG.2.2.29595.03364,   by Aguirre {\it et al.}, 
 and   the 2005 {\it CERN Courier} article, by  L. Maiani,  
  \url{https://cerncourier.com/a/viewpoint-latin-america-comes-to-cern/}.}} 

{While history cannot repeat itself, it can be inspirational  to revisit   
 the major role played by the science of particle accelerators in the last century,  by recounting how   the construction of the Frascati electron synchrotron  
  in Italy spear-headed 
   the country  scientific recovery after WWII, at the same time  underlying the contemporary institution of the intergovernmental Conseil Europe\'en pour la Recherche  Nucl\'eaire (CERN). }

Particle accelerators were developed in the first half of the last century to probe the structure of matter. 
{From  linear accelerators to   synchrotrons, 
 the  history  of their development and their modern relevance  to  society's needs  is well known, less known is how an electron synchrotron led to the development of a whole new branch of particle physics, when, in Italy, in 1960,  the Frascati electron synchrotron was 
instrumental to the birth of   the first ever electron positron collider, {named  AdA,  an Italian acronym for {\it Anello di Accumulazione}, a Storage Ring in English}.
To host a large enterprise for a new generation particle accelerator, such as the Frascati synchrotron, a  national laboratory, had been built between 1953 and 1959,    fostering  the gathering of a pool of scientists, technicians and  engineers, trained in the post-war electronics and vacuum technology,  made possible by the creation of an intergovernmental institution in Europe  such as 
CERN,  and  the  international transfer of knowledge  between Europe and the United States, which fostered the European reconstruction.
 Thus new ideas, such as  the  electron-positron collision experiment,  which  Bruno Touschek  \citep{Amaldi:1981be} proposed \citep{Bernardini:1960osh}, could arise 
  and their realization was made  possible  \citep{Bernardini:1963dyf,Bernardini:1964lqa}.}
 
\begin{figure}
 \centering
 \includegraphics[scale=0.24]{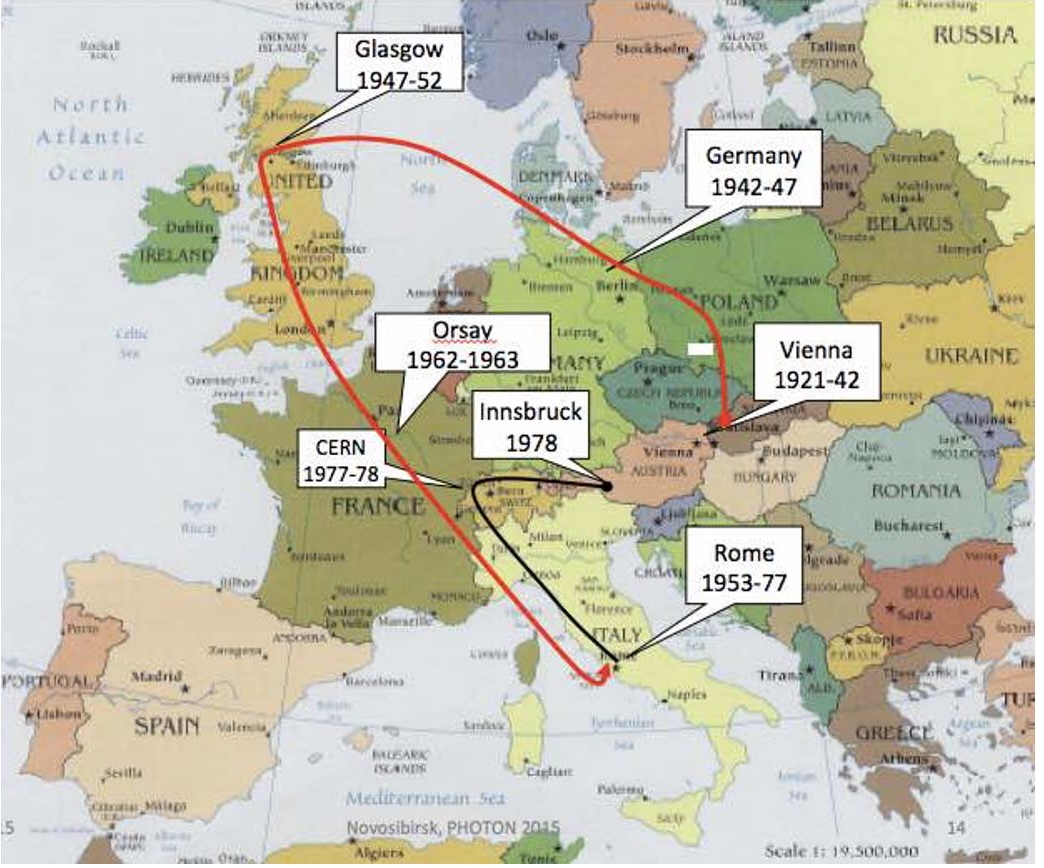}
 \includegraphics[scale=0.31]{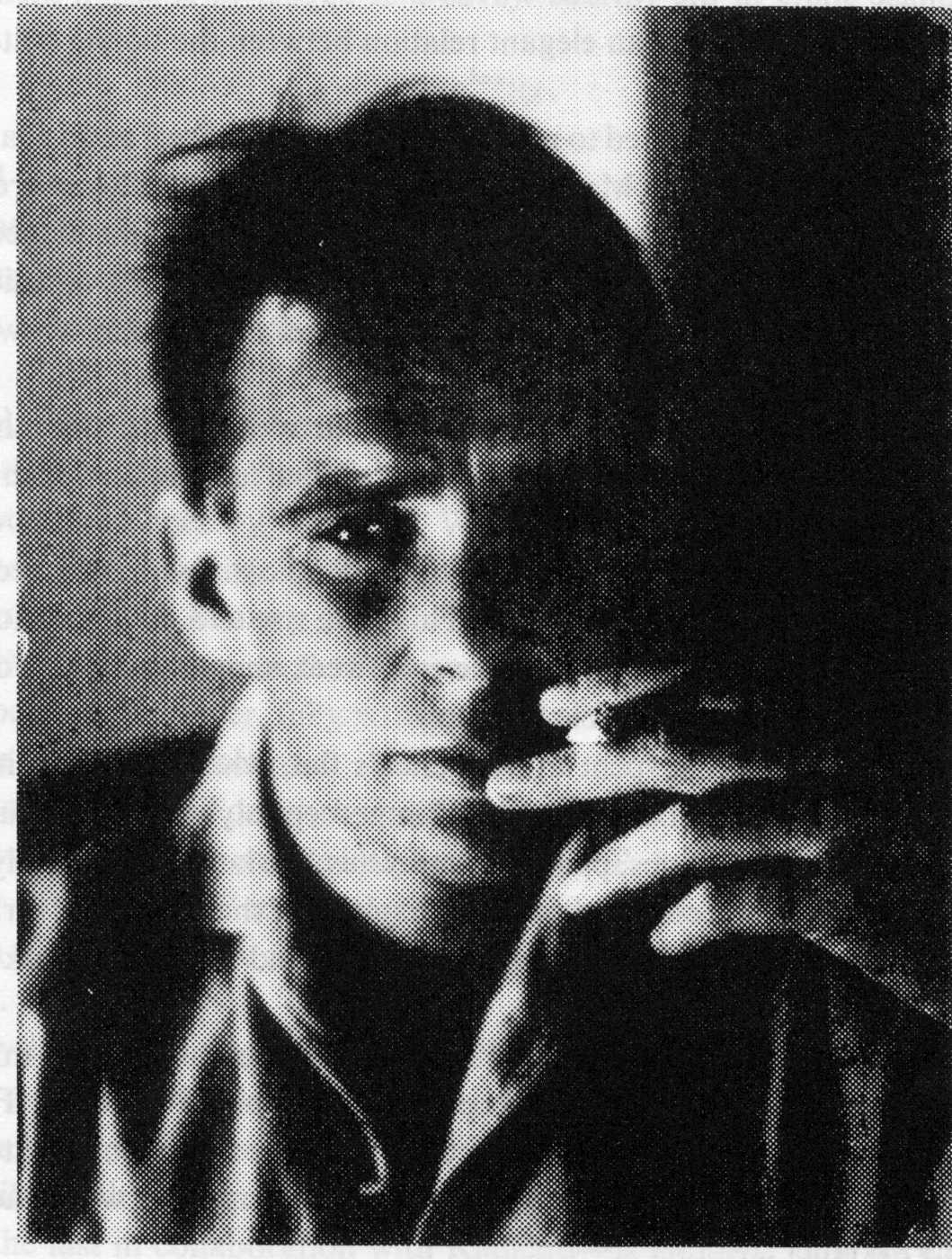}
  \includegraphics[scale=0.061]{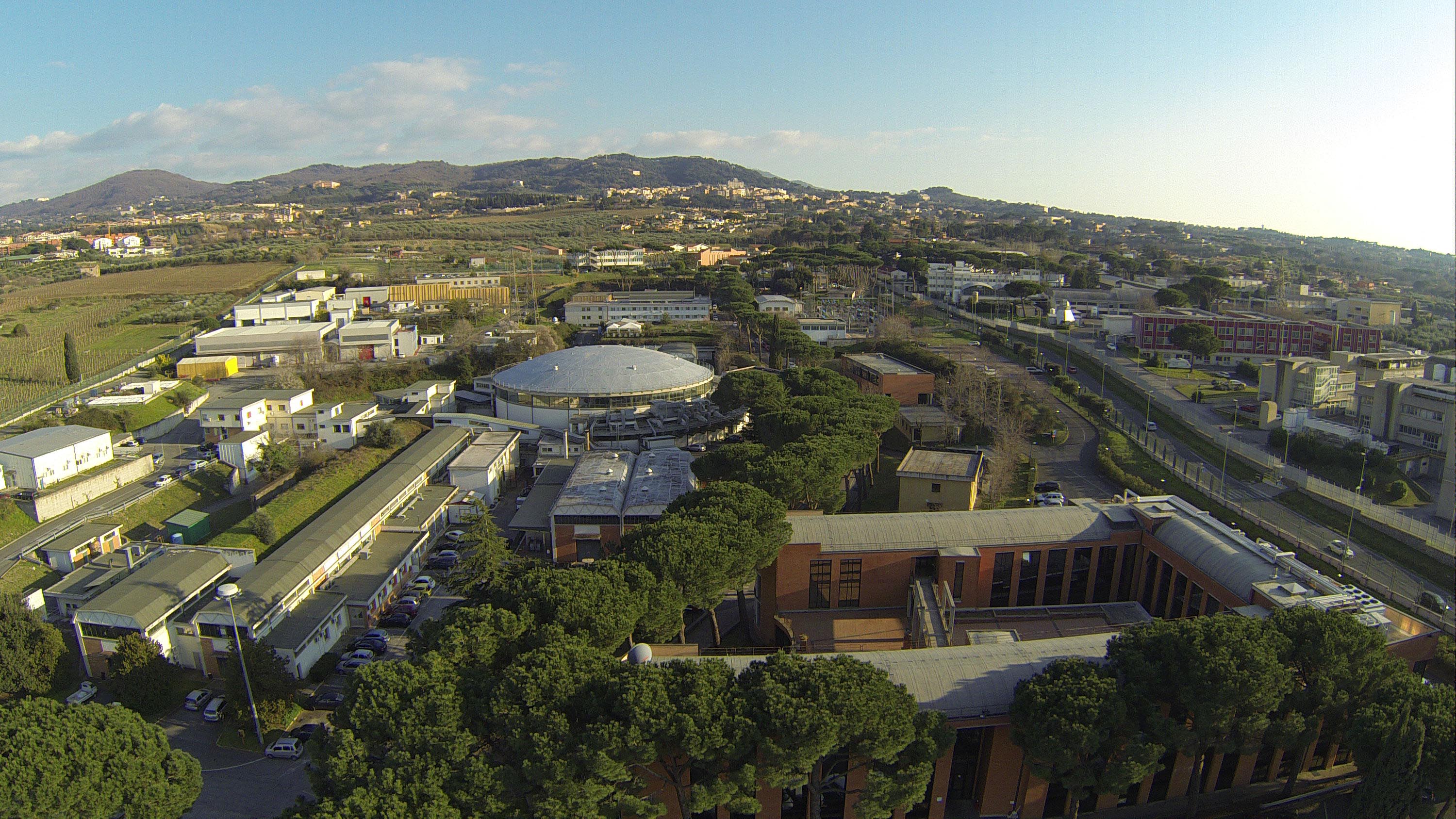}
 \caption{{Left panel:}   {image} depicting Bruno Touschek's life through Europe {in G.P.'s  drawing on a  1941 public domain  map}. {Center panel:}  Bruno Touschek in 1955  in Rome, courtesy of  CERN Documentation Services. {Right panel:}  contemporary view of the Frascati National Laboratories with the ADONE building, where electrons and positrons, accelerated to 2.8 GeV center of mass (c.m.)  energy,   observed an anomalous production of hadronic particles in 1969, 
 with  the old synchrotron building,  at right across  the {road},
  Italy, \copyright \ INFN-LNF image.}
 \label{Fig1-BTpower}
 \end{figure}

Now, more than sixty years since the first laboratory observation of center-of-mass electron positron collisions { in AdA}
  the European Strategy Group for particle physics has  proposed that the  next-generation particle accelerator   be a 
 circular collider, FCC-ee,  with electron-positron collisions {as}   the starting step for a future  hadron-hadron collider.\footnote{For details, such as the 91 Km length, more than   ten   thousand times longer that  AdA, see \url{https://home.cern/the-cern-council-decided-to-update-the-european-strategy-for-particle-physics/} and {\url{ https://physics.aps.org/articles/v19/127} }}  
  What began  as an ``experiment worth having", in the words of  Bruno Touschek {\citep[p.319]{Pancheri:2022nzs}}, the Austrian born physicist who proposed it in 1960, has now become the   aim of the latest generation of physicists, making worth knowing  the story of how that ``experiment" (with the Frascati synchrotron)  was envisaged and realized. How that happened is an extraordinary combination of circumstances, which includes the reconstruction of European physics through the institution of a regional center {such as}   CERN,  and {the creation of} two national laboratories, the \LAL \  d' Orsay, in France, and the  Frascati National Laboratories in Italy. But, perhaps, it {wouldn't}  have happened then and there,  and it  could have happened later and in a different part of the world, for instance in the USSR, in the Nuclear Physics Laboratory of Novosibirsk, beyond the Ural mountains, and who knows how different things would have been. What  certainly made the difference was the existence of { an intergovernmental collaboration flanked  by  national laboratories, which harboured and nourished }   the vision and determination of a theoretical physicist, Bruno Touschek, whose life crossed Europe,  in space and time, Fig.~\ref{Fig1-BTpower}. 

   To  understand how many different stories had to be woven together for a new field of research to be born, such as happened with electron positron physics in the 1960s, I will  start, in Sect.~\ref{sec:colliders}, with a brief  outline of  the development of particle accelerators, from the earliest attempts,  to {synchrotrons and the rise of collider ideas
  in the USA.}  In Sect.~\ref{sec:CERN-LAL-LNF}, I will  describe how the creation of CERN gave rise to the birth of two national laboratories in Europe    and  summarize the  birth  of AdA   in Sec.~\ref{sec:birthofada}.  
{To picture the creation of AdA in the personal and formative experience of its creator, in Sect.~\ref{sec:BT} I shall give a brief introduction to Bruno Touschek's life  after  he left  his native Vienna in 1942 \citep{Amaldi:1981be}, \citep{Greco:2004np}, \citep{Pancheri:2022nzs}   and \citep{Bonolis:2023dle}, up to when  he joined  the Physics Institute in the University of Rome in January 1953.}
  
    {Highlights of the impact of Touschek's talk at  the  CERN  1961 Conference on Theoretical Physics  \citep{Bell:1961gi} and  the international Franco-Italian collaboration which allowed to prove the feasibiity of  electron-positron colliders will be presented  in  Sec.~\ref{sec:ada-Europe}.}
  
 {\bf {Sources}}
{An extensive  list of both primary and secondary sources  for the events described in this article can be found in \citep{Pancheri:2022nzs}. Specific events cited  here 
 include   journal articles and  books, in addition to documents deposited in university or research center archives, as well as privately held material such as personal letters. Documents from the following institutional archives have been used: \RSUPD \ for Edoardo Amaldi and Bruno Touschek papers and photographs, INFN-LNF Library and LNF Images Collection,  CERN Documentation Services,  Orsay \LAL \ Archives, and Vienna Jewish Archives. Copy of  letters from \LAL \ Archives were sent to this author by e-mail in 2011 by Prof. \JH, emeritus Professor at Orsay.  Among privately acquired document are copies of letters exchanged between Bruno Touschek and his family, photographed  with Touschek family permission in 2009 by  this author and L. Bonolis, with the original letters  privately  held by Touschek's son.  Interviews used for this  text took place  in 2003 in Frascati National Laboratories, in Paris in 2013  at the privately held residence of Prof. Maurice L\'evy,  and in Rome in 2010 and 2013, at Prof. Giorgio Salvini's home,  and  were partially used in the
 docu-film  {\it Touschek with AdA in Orsay} by E. Agapito, L. Bonolis and G.P., produced by INFN and Media Scienza.    The original reels are with the author Enrico Agapito and one copy with this author. Plans are ongoing to deposit all personal digital material in  \RSUPD.} 
 \section{What is a collider and why antimatter}
  \label{sec:colliders}

     \begin{figure}
 \centering
    \includegraphics[scale=0.3]{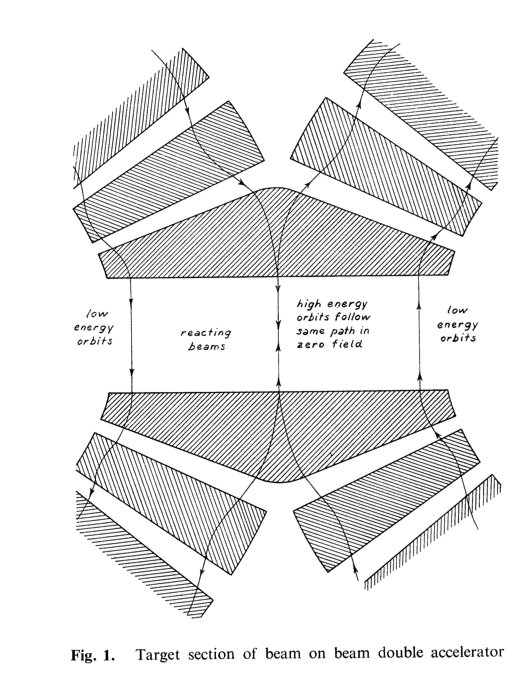}  \hspace{0.3 cm}
  \includegraphics[scale=0.3]{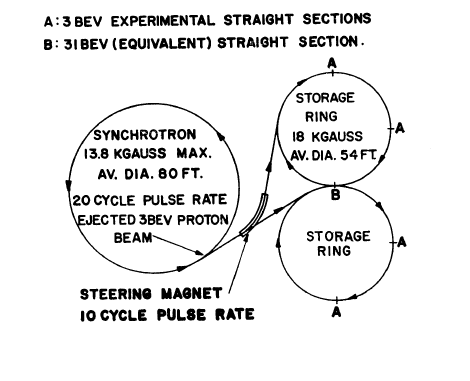}
 \includegraphics[scale=0.47]{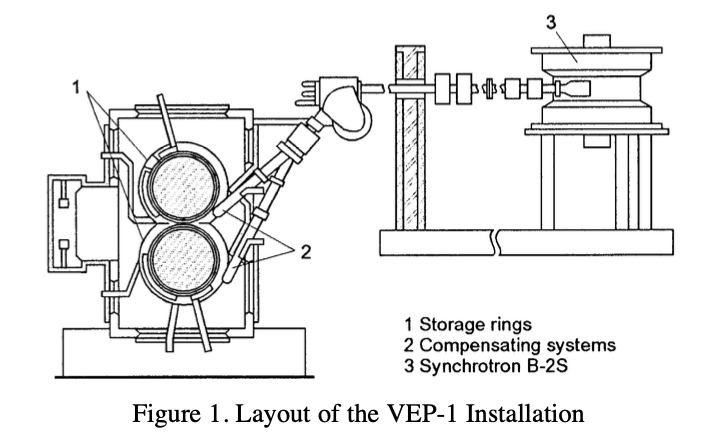}
 \includegraphics[scale=0.155]{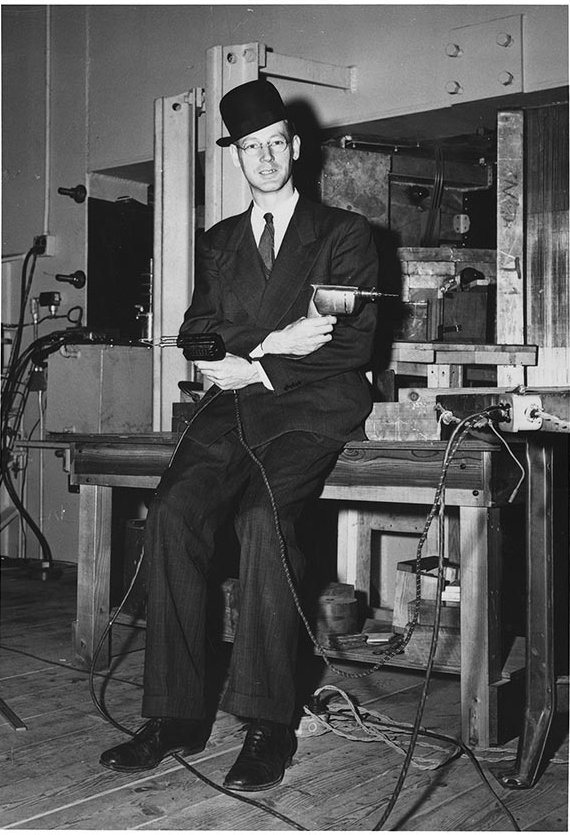}
  \includegraphics[scale=0.25]{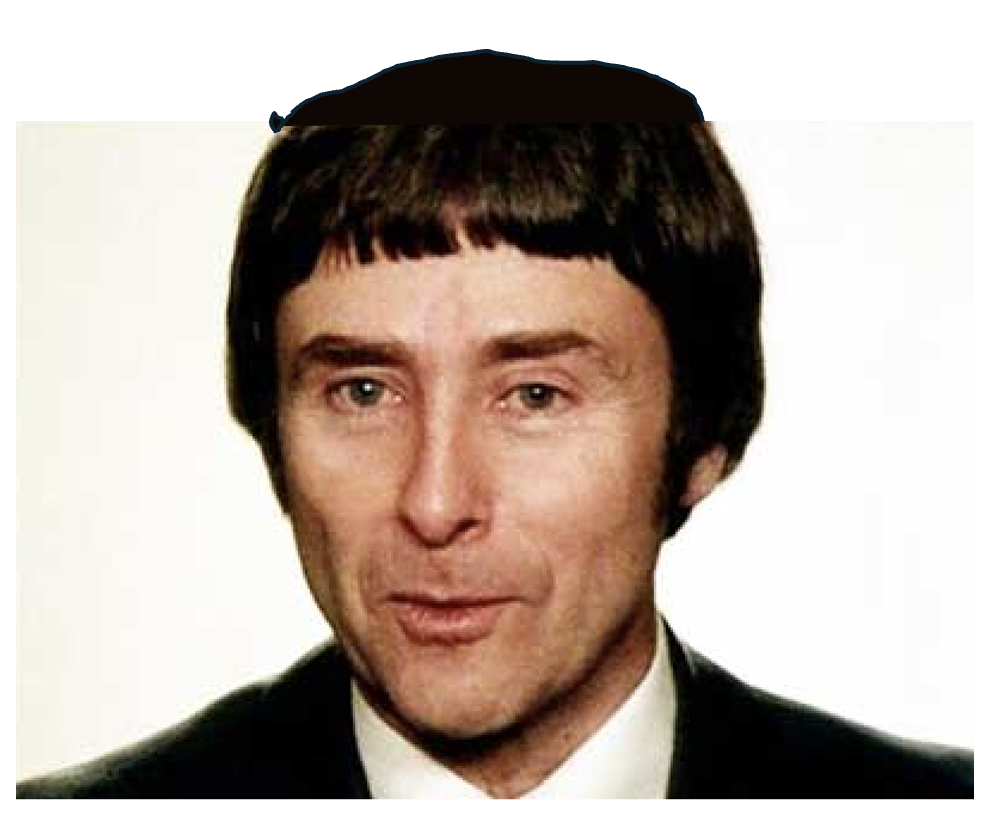}
    \includegraphics[scale=0.24]{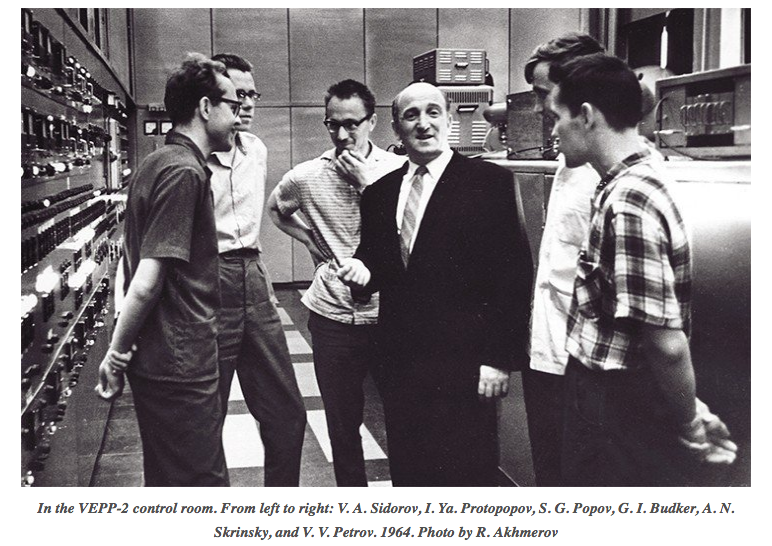} 
 \caption{{Top level}:  at lefy  the  schemes for electron-electron or proton-proton collisions   proposed by D. Kerst (left),  by G. O'Neill at center, from  the Proceedings of the {\it 1956 CERN Symposium on High-Energy Accelerators and Pion Physics}, Geneva, CDS,   and, at right,  the scheme for VEP-1,  built by  G.I. Budker,  from a recollection by A. Skrinsky in 1996. {Bottom panel}: the  respective proposers of the above schemes,  D. Kerst (left) , photograph courtesy of University of Illinois,   G. O'Neill (center), public domain,    and  G.I. Budker  and his collaborators  (right) in the VEPP-2 control room, from A. Skrinsky's  contribution, to the centenary volume  {\it Budker in Four Perspectives }   \url{https://scfh.ru/en/papers/budker-in-four-perspectives/}.}
\label{fig:KerstOneillBudkerscheme}
 \end{figure}
  A particle collider realizes the intuitive  idea of center of mass collisions as a means of reaching higher energies and make easier to break up  nuclei and their components, to access their inner structure. 
  
  The science of particle accelerators was developed in the last century, starting with a  linear accelerator for electrons \citep{Ising:1924, Wideroe:1927lju}   then moving to circular structures, such as cyclotrons \citep{Lawrence:1931cb},
  betatrons \citep{Kerst:1941zz,PhysRev.60.53} and synchrotrons
   \citep{McMillan:1945zz,PhysRev.69.244}
   all of them consisting of accelerating a single beam of charged elementary particles, such as electrons or protons, 
  impinging on a target of nuclear material  and observing the scattered broken content. Large laboratories were built to house the  accelerator facilities, which became bigger and bigger to allow for higher  particle accelerations and higher energies hitting the target. 
  
  In the mid 1950s the idea of  exploiting the kinematic advantage of center of mass collisions, first suggested by \RW \  during WW-II \citep{Waloschek:1994qp}, inspired US \citep{Kerst:1956whq,ONeill:1956iga}
   and USSR \citep{Baier:2006ye}
scientists to develop accelerators where beams of electrons or protons moving at the same speed in  opposite directions would clash against each other. Projects with electrons against electrons started in the USA,  at SLAC,  the Stanford Linear Accelerator Center \citep{Barber:1959vg}, and in the USSR between Moscow and the new laboratory of the Siberian Branch of the Russian Academy, in Novosibirsk \citep{Skrinsky:1995kh},  
at the end of the  1950s, Fig.~\ref{fig:KerstOneillBudkerscheme}. 
  
{But, as described next,  it was in Europe that the spark of future accelerators was ignited, as post-war reconciliation had spurred new collaborations fostered by
intergovernment  exchanges.}

\section{{CERN and the creation of national laboratories in Europe} }
\label{sec:CERN-LAL-LNF}
\begin{figure}
\centering
\includegraphics[scale=0.2]{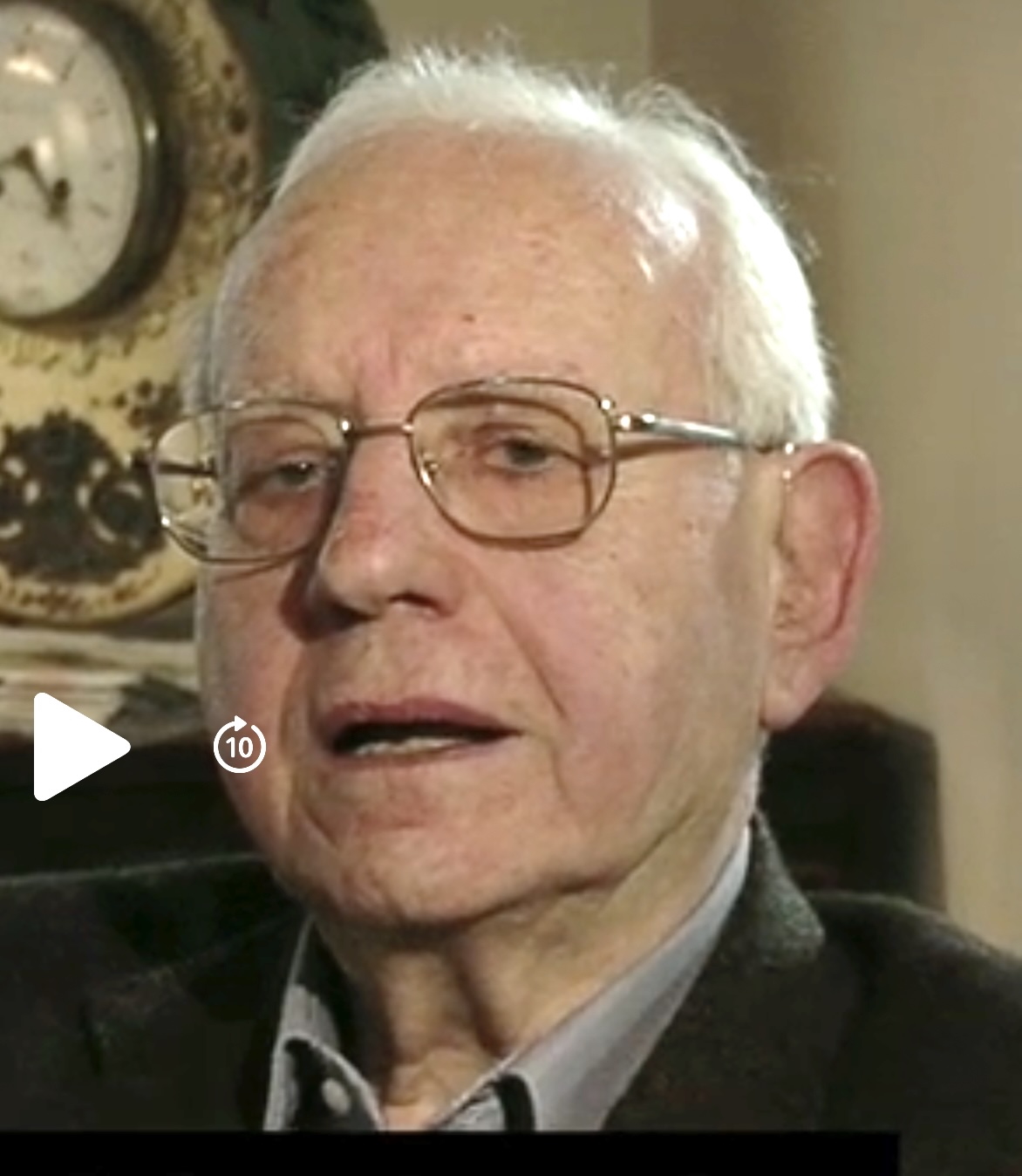}
\includegraphics[scale=0.204]{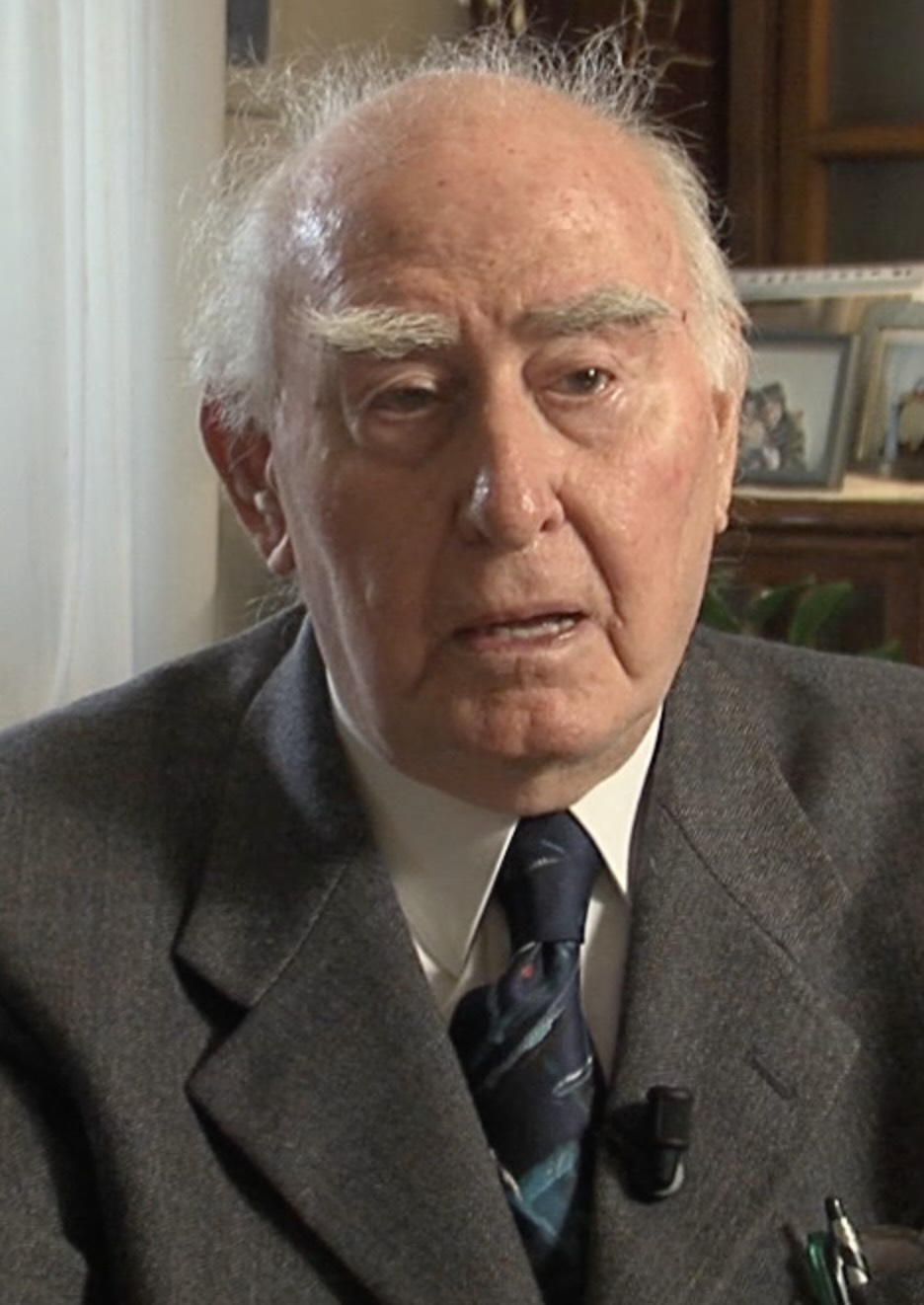}
\includegraphics[scale=0.501]{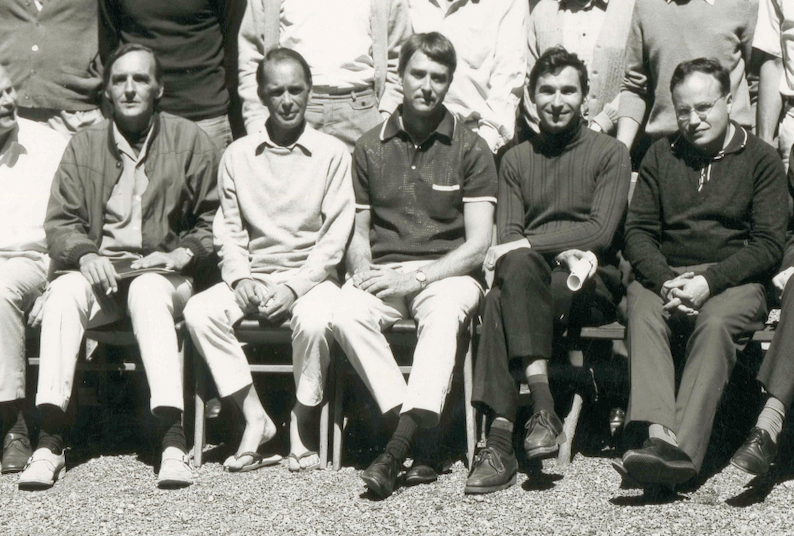}
\caption{{Left panel: Maurice L\'evy, from May 2013 interview for the docu-film {\it Touschek with AdA in Orsay}, \copyright \ E. Agapito and INFN. Center  panel: Giorgio Salvini,  2010 in the video {\it Salvini Giorgio,  Considerazioni a 90 anni}, \copyright \ E. Agapito and INFN. Right panel:  extract from a group photo with Matthew Sands at left, next to B. Touschek, G. O'Neill, J. Ha\"issinski and V.N. Baier, 46th International School of Physics "Enrico Fermi" : Physics with intersecting storage rings, 16 - 26 Jun 1969, Varenna, Italy, Societa' Italiana di Fisica, reproduced with permission.}}
\label{fig:levysalviniamaldi}
\end{figure}
{The series of events sketched in this article is based on oral and written contributions,  
photographic and video recordings, which started  with Amaldi's biography of Touschek \citep{Amaldi:1981be} and continued   in 1987 with the
 {\it Bruno Touschek Memorial Lectures}, held in Frascati  to
commemorate Bruno Touschek's untimely demise in 1978  \citep{Greco:2004np}. 
Among the testimonies gathered  since then, there are  video interviews taken
  for educational purposes in 2003, 2010 and 2013,  realized   with the support of  INFN,  the Frascati National Laboratories, and Orsay \LAL. Complementing  these interviews
 with  the 1987  AIP   interview with the American physicist Matthew Sands, 
  allows a clear picture to emerge about  the post-war rise of particle accelerators in Europe.}

{The  birth of national laboratories in Europe in the 1950s was a natural consequence of the creation of CERN, as recalled below  in excerpts transcribed from  video   interviews 
 taken    in 2013,   in Paris with  the eminent French theoretical physicist Maurice L\'evy, at left  in Fig.~\ref{fig:levysalviniamaldi},  and in Rome with the first director of the Frascati Laboratories, Giorgio Salvini, {\it ibidem} at center,   for  the docu-film {\it Touschek with AdA in Orsay} by E. Agapito, L. Bonolis and G.P., \url{https://w3.lnf.infn.it/ada-anello-di-accumulazione/}, \copyright \ Agapito Mediascienza. }

{\begin{description} \item[Maurice L\'evy (Paris, May 2013)]
I came from the United States in 1952 at the \'Ecole Normale Sup\'erieure [in Paris] that at the time was directed by Yves Rocard,  the father of the well known politician Michel Rocard,  and he  asked me to start a group  of theoretical  physicists. [\dots] In 1953, we started to have a discussion about the creation of particle machines. The reason was that the decision to create CERN was taken in 1952. It was a group of physicist, Pierre Auger from France, Edoardo Amaldi from Italy, Harry Massey from England, Heisenberg from Germany, they decided to create CERN, the Laboratory, which was approved by the governments,  and this created immediately a problem, for us and I believe for the rest of Europe, in the sense that we did not have enough experimentalists to work both at CERN and in Paris,  in France, in general. So it was decided by discussions   with Francis Perrin, who was at the time the Haut Commissaire pour l'\'Energie Atomique, and Yves Rocard, to start the building of two machines, one was a proton circular machine  to be built in Saclay by the Atomic Energy Commission, and  it was called Saturne, and the second one  was supposed to be an electron machine.   For the electron machine there were two possibilities, one was to create  a circular machine of the type  Bob Wilson   had built in Cornell University, the other  was to build a linear accelerator of the type at the [Stanford] Linear [Accelerator] Center. [\dots The] final decision was taken  to build a linear accelerator in Paris. [\dots] In Paris the question was 
 we could not build a linear accelerator at the \'Ecole Normale which  did not have the space and because of that, between '55 and '60, we were developing  an new campus in Orsay, where we had all the space we needed and, because of that, the building of the LAL [\LAL] and building of Orsay went together.
\end{description} }

{ In Italy, the decision to build a large laboratory outside Rome was taken in 1952 by  Edoardo Amaldi,  CERN first Secretary General, and INFN president Gilberto Bernardini \citep{Battimelli:2001aa}. The task of building the laboratory and the machine went   to Giorgio Salvini, who  had done his physics dissertation on the betatron in 1942 and, after the war, had spent two years at Princeton, doing cosmic ray detector research, returning  to Italy, in 1952.  The development  of  decisions in Italy is discussed by Giorgio Salvini,  in a  June 2013 video-interview. 
 \begin{description} 
 \item[Giorgio Salvini (Rome, June 2013, in Italian, author's translation)]
In '47-'48 Edoardo Amaldi, Gilberto Bernardini and others were asking what could they do for Italy's renaissance. These great personalities got the idea to build a new laboratory and to place there an accelerator. 
In the meanwhile the decision for a French machine had been taken, and in Germany, which had been totally destroyed,  the [industrial and scientific] machine was restarting. 
[\dots]  In 1952, back to Italy from the United States, the prospect to build a new laboratory arose, and I was asked to be the director of the new laboratory. From what and where?
At the end the choice fell to Frascati.\footnote{The choice of location came from the government, and through  connections between the Ministry of Industry,  the mayor of Frascati and other locally  important political players 
 \citep[p.39]{Valente:2007}.} 
 Frascati? I had never heard of it. [\dots] A friend  warned me of the risk  to be stuck all my life in a nowhere land.
\end{description}}
{In an AIP interview by Finn Aaserud,\footnote{Santa Cruz, California
Transcript version date: December 18, 2024
\url{https://doi.org/10.1063/nbla.lhoa.awxy} May 4 and 5, 1987.} the American physicist Mathew Sands, in the right panel of   Fig.~\ref{fig:levysalviniamaldi},  described how the building of the Frascati synchrotron changed Italy's prospects from those of a country which had lost the war into  the successful protagonist  of  a new scientific renaissance:  }
{\begin{description}
\item[Matthew Sands (May 4 and 5, 1987)]
[In 1952, I went for sabbatical  from Caltech to Europe, to Paris and  Rome, where \dots]  Oh, it was postwar poverty. I mean, they had no money. They still had no money for
research. They had managed to scrounge a few microscopes to work with emulsions.
They would get surplus electronic radio gear from the army to steal parts from to make
counting circuits for cosmic rays. 
However, it seemed to be the breaking point, I mean the transition point. It was 1952,
and this was the time when things were going to change. CERN was just getting
organized. The government was proposing that it could find money to support research.
\end{description}
 In April of 1968 Sands  returned  to work for six months in Italy, where he found  Italian physicists to  now be in the forefront of accelerator science, indeed, as he said in the AIP interview, 
 \begin{description}
\item[Matthew Sands (May 4 and 5, 1987)] \dots they were now the world leaders in this field.
\end{description}
The transformation, which would make   Italy   a major  player in accelerator physics development,
    had taken place with a  new transition point  easily identified at the end of 1959, when the Frascati laboratories had been built and an electron synchrotron had started operating, prompting the search for new ideas, as we shall  see next.}

{Sand's words are echoed  by Giorgio Salvini, interviewed  in 2010, in preparation of celebrations for   the  150 years since Italy's unification in 1870.\footnote{The video-interviews was taken in G. Salvini personal residence in Rome,  by E. Agapito,  on the occasion of Salvini's 90th birthday, \url{https://lamediateca.infn.it/mediateca/view.php?v=213}.}
Ninety year old at the time, he had been Professor of Experimental Physics at the University of Rome   for almost sixty years,    President of the National  Accademia dei Lincei and Italian Minister for University and Scientific and Technological Research.
\begin{description}
 \item[Giorgio Salvini(Rome, May 2010, in Italian, author's translation)]
 It is now 150 years [since the unification of Italy] of which i have only lived through 70 of them. 
  I can testify to  the enormous contribution to Italy's unification brought by science and, in particular, by the science in which I had the good luck to operate.  
\end{description}}

\section{{The birth of electron-positron colliders}}
\label{sec:birthofada}
{As   1959 ended, seven years after the events recollected by L\`evy  and Salvini},  in  Europe there were  {three new laboratories, and three new elementary particle accelerators}, a new generation proton synchrotron, the PS at CERN \citep{Brianti:1997iq}, 
a powerful linear accelerator at  the \LAL \ in Orsay \citep{Marin:2009} in France, and an electron synchrotron at Frascati \citep{Salvini:1962aa} 
in Italy, which, for a while, had the highest energy ever reached in a laboratory. European physicists were discussing and meeting, travelling to conferences, and exchanging ideas with both Russian and American counterparts, often at CERN, {where accelerator physicists were looking for new ways to reach even higher energies and plans, for a  conference to discuss the future of particle physics, were being laid by a group of four theoretical physicists. It was  at  such a conference, held in June 1961 \citep{Bell:1961gi}, that colliding beam physics came to the fore, when Touschek announced that Frascati had built an  electron positron  storage ring, where matter would clash against  anti-matter.}  But how did the idea to use matter-antimatter arise? And  how feasible could it be? The turn is now for theoretical physics to join the field, as  a powerful concept,  invariance of an elementary  particle state under transformations of Charge, Parity reflection  and Time reversal (CPT),  was formulated into a theorem  between 1953 and 1958 by a number of scientists  \citep{Blum:2022eol},  among whom there was  Wolfgang Pauli,  the discoverer of the exclusion principle.\footnote{W. Pauli was  recipient of the 1945 Nobel Prize in Physics ``for the discovery of the Exclusion Principle, also called the Pauli Principle", \url{https://www.nobelprize.org/prizes/physics/1945/summary}/. Pauli's contribution to the formulation of the CPT theorem is included in W. Pauli, {\it Exclusion principle, Lorentz group and reflection of space-time and
charge}, in: {\it Niels Bohr and the Development of Physics}, pp. 30–51 (1955),
 https://cds.cern.ch/record/96173/files/CERN-ARCH-PMC-05-029.pdf.}   During these years, there were frequent exchanges between Pauli     and Bruno Touschek  who had joined the Institute of Physics in Rome in 1953, after having held a Nuffield Lecturer position at the University of Glasgow \citep[Chap. 7]{Pancheri:2022nzs}.  Thus Touschek learned about the CPT theorem, while it was in the making,\footnote{{Exchange of letters between Bruno Touschek and other proposers of the CPT theorem, such as Gerhard L\"uders, are held in Touschek Papers in  \RSUPD.}}  and, when the time came, he proposed to the Frascati scientists  to  realize it in a laboratory, by making matter-antimatter collisions, with particles of opposite charge, moving in opposite directions.  
 \begin{figure}
 \centering
 \includegraphics[scale=1.04]{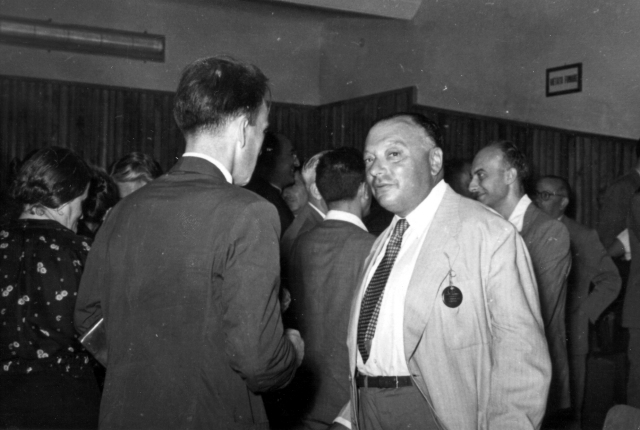}
 \includegraphics[scale=0.172]{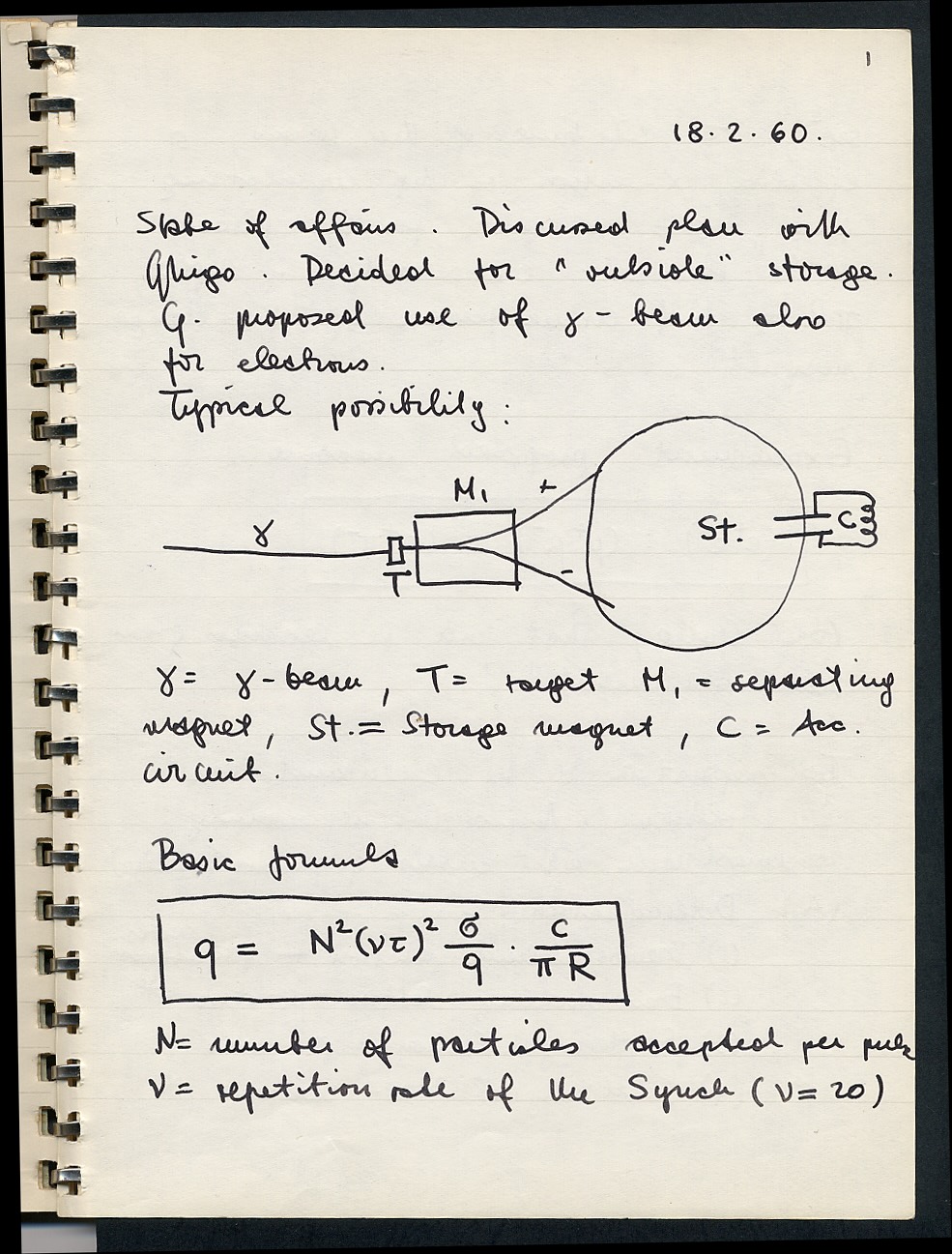}
 \includegraphics[scale=0.16]{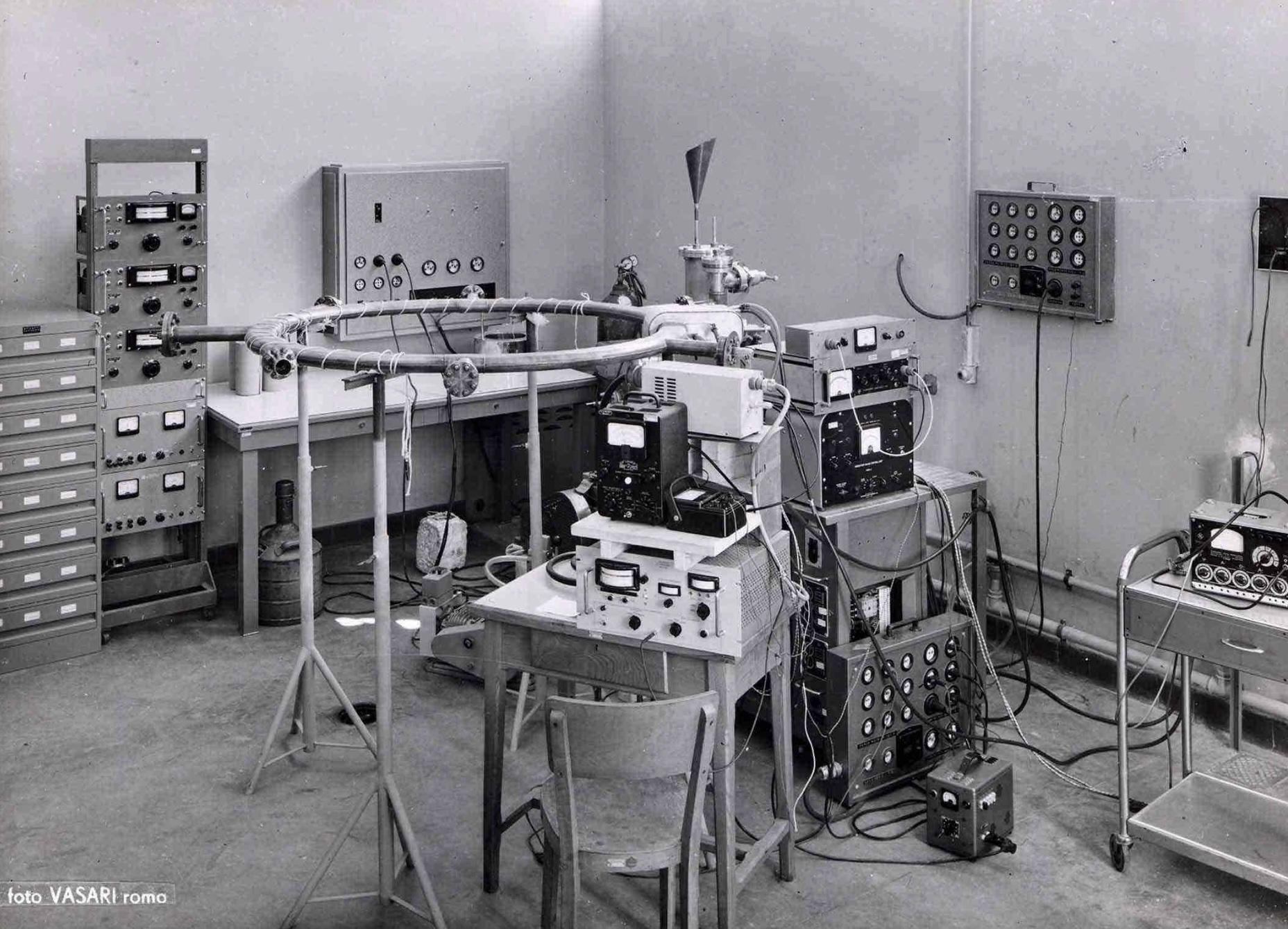}
 \caption{{Left panel:} Bruno Touschek (back to the camera) with Wolfgang Pauli, in 1953 in Cagliari, during the National Congress of the Italian Physical Society, courtesy of CERN Documentation Service. {Center panel:}   the first draft of AdA's scheme prepared in February 1960 by Bruno Touschek, \copyright \ Touschek family,  on the day which followed his proposal for an experiment  with the Frascati synchrotron. {Right panel:}   the steel  doughnut for electrons and positrons to circulate together, under preparation at the Frascati Laboratory in 1960, \copyright \ LNF-INF images. 
 }
 \label{fig:BTPauli-AdaBt-ciambella}
 \end{figure}

\noindent {\bf Testimonies  from the past}\\
\RW\   \cite[p. 81-82]{Waloschek:1994qp}:
{\it An important event in my Hamburg period  happened during the Autumn 1943 [\dots] which started me thinking that [\dots]  when a target particle (at rest) is bombarded, [\dots] only a relatively small portion of the accelerated particle's energy is used to actually split or destroy the colliding particles. However when the collision is frontal most of the available \dots energy can be exploited.  [\dots] I spoke with Touschek about my ideas and he said they were rather obvious \dots}\\
Bruno Touschek (February 1960):\footnote{{\BT,  {\it On the Storage Ring}, incomplete typescript,  \RSUPD, Touschek Papers,  3.86.1 in \citep{Giovanni-Battimelli:1989aa}.}} 
\begin{tt}
The first suggestion to use crossed beams I have heard during the war [was]
 from \W \dots
 \end{tt}\\
 Nicola Cabibbo  \citep{Cabibbo:1997aa}:
 {\it I still recall vividly the seminar in Rome  in late '59 by [W.] R. Panofsky. It was after the seminar that Bruno Touschek come up with the remark that  an $e^+e^-$ machine could be realized in a single ring, ``because of the CPT theorem".}\\
Bruno Touschek (1974):\footnote{{Bruno Touschek, {\it A brief outline of the story of AdA}, excerpts from a talk delivered at  the Italian Accademia dei Lincei 24-05-74,  \RSUPD,  Touschek Papers 3.92.5 in \citep{Giovanni-Battimelli:1989aa}.}}
    \begin{tt}
 The challenge of course consists in having the first machine in which particles which do not naturally live in the world which surround us can be kept and conserved.
  \end{tt}

  The path to realize such an innovative idea as matter-antimatter collisions in a laboratory setting, can be traced to the extraordinary, and often tragic, life of Bruno Touschek, the scientist who proposed AdA  in February 1960,  and proved its feasibility. 
  
\section{{Bruno-Touschek (1921-1978) from Vienna  to Rome}}
\label{sec:BT}
 Bruno Touschek\footnote{{The  biographical material in this section was  first collected by E. Amaldi directly from Bruno Touschek in February 1978, and was complemented by archival documents received by this author from the Vienna Jewish Archives. Of particular interest for the period covered in this section, have also been the letters sent by Touschek to his father from 1939 to 1953, so far unpublished, with excerpts in \citep{Pancheri:2022nzs}. }} was born in Vienna in 1921, in a family immersed in the artistic movement of the Viennese Secession movement. His mother was Jewish, his father a catholic and a Major in the  Austrian Army. As a young boy, Bruno suffered two successive tragic losses, his mother in 1931, and his maternal uncle, Oskar Weltmann, a doctor and a painter  he very much admired, by suicide  in 1934. After losing two of her children, his maternal grand mother, Josephine Weltmann, decided to move to Rome and live with her only surviving daughter, Adele, nicknamed Ada, who was married to an Italian industrialist.  Thus, when the annexation of Austria to Germany in March 1938 destroyed the life and    well being of Vienna Jews, Bruno's father thought wise to send Bruno to study in Rome, where anti-semitism, {as perceived from Vienna},  was not yet rampant and he could attend engineering  courses  at the University, while living with his maternal aunt and grandmother.  In Rome, one of the courses he attended was 
  on mathematical physics,  taught to both  physics and engineering students, by a famous professor, Francesco Severi. Bruno had not yet passed his final high school exams, the so called {\it Matura},  and  the course was probably more advanced than what Bruno, only 17 years old, could successfully study, but his
  fascination with theoretical physics may have started here, as he later wrote
   to have attended {Severi's course}  with \begin{tt} ``more enthusiasm than success"\end{tt}.\footnote{{See the draft of Touschek's   last CV, undated,  but probably  prepared  in late February 1978,  Amaldi Papers in \RSUPD.}}

  The project to stay in Rome did not materialize, since   the life of Italian Jews  under Mussolini's government,  
  { took a turn for the worst} after Hitler's visit to Rome,  followed by anti-semitic laws  promulgated in October and signed by the King of Italy in December 1938 
  \citep{Violini:2006}. 
  Bruno returned to Vienna, as did  his grandmother. 
  
In 1939, as  the position of  Jews in Vienna was further deteriorating,  Bruno made an attempt to  emigrate to the UK, and study  Chemistry at the University of  Manchester, a city where a substantial Jewish community had been established since the previous century.
For reasons unknown, Bruno did not succeed in going  to the UK,   and in September 1939   enrolled to study physics at the  University of Vienna, where  a  great school in mathematical physics 
included   such professors as 
 Hans Thirring, later a friend and  mentor of Touschek's. Following a suggestion by another of his teachers, Bruno's road as a theoretical physicist took its definitive direction when he  was  encouraged  to read Arnold Sommerfeld's treatise {\it Atombau und Spektrallinien} and follow this up with a correspondence with the great scientist in late December 1941 \citep[p. 36-37]{Pancheri:2022nzs}. Sommerfeld encouraged Touschek to move to Germany in order to continue his studies and gave  Bruno access to his vast network of former students and pupils, many of them professors in the German universities, such as Paul Harteck in Hamburg, Max von Laue and Werner Heisenberg in Berlin.\footnote{ Von Laue and Heisenberg were both  Nobel Prize recipients,  Von Laue in 1914  ``for his discovery of the diffraction of X-rays by crystals" and Heisenberg in 1932 ``for the creation of quantum mechanics, the application of which has, inter alia, led to the discovery of the allotropic forms of hydrogen”, \url{https://www.nobelprize.org}. } In Germany for 5 years, from  1942 to March 1947, Bruno learnt theoretical physics from some of the  scientists who had developed it during the pre-war years  or were still developing it, as was  the case with Heisenberg. After the war, Touschek   become  his assistant, in \Gott, where  German science was being reconstructed under Heisenberg's direction. 
 \begin{figure}
 \centering
   \centering
\includegraphics[scale=0.132]{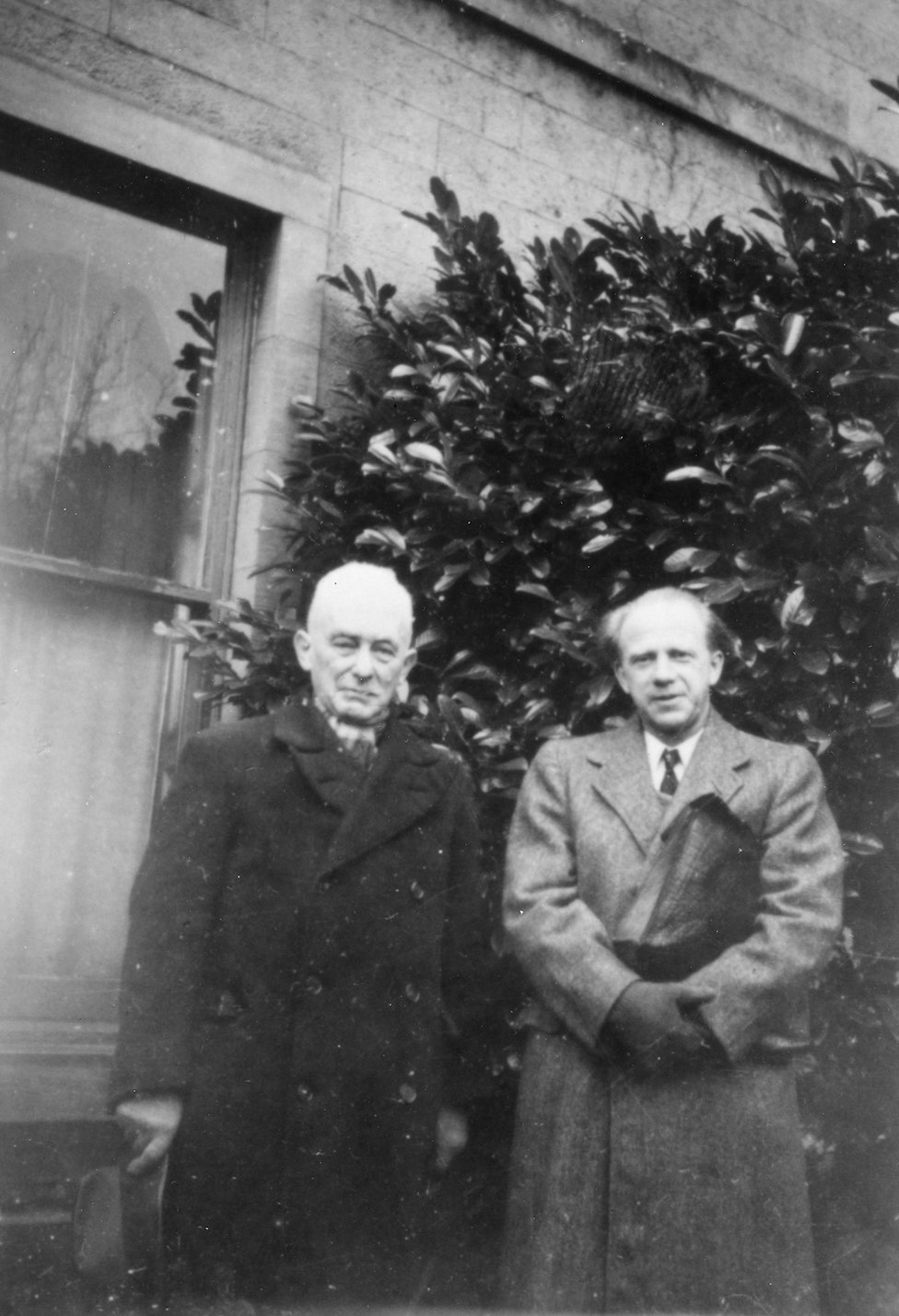}
  \includegraphics[scale=0.298]{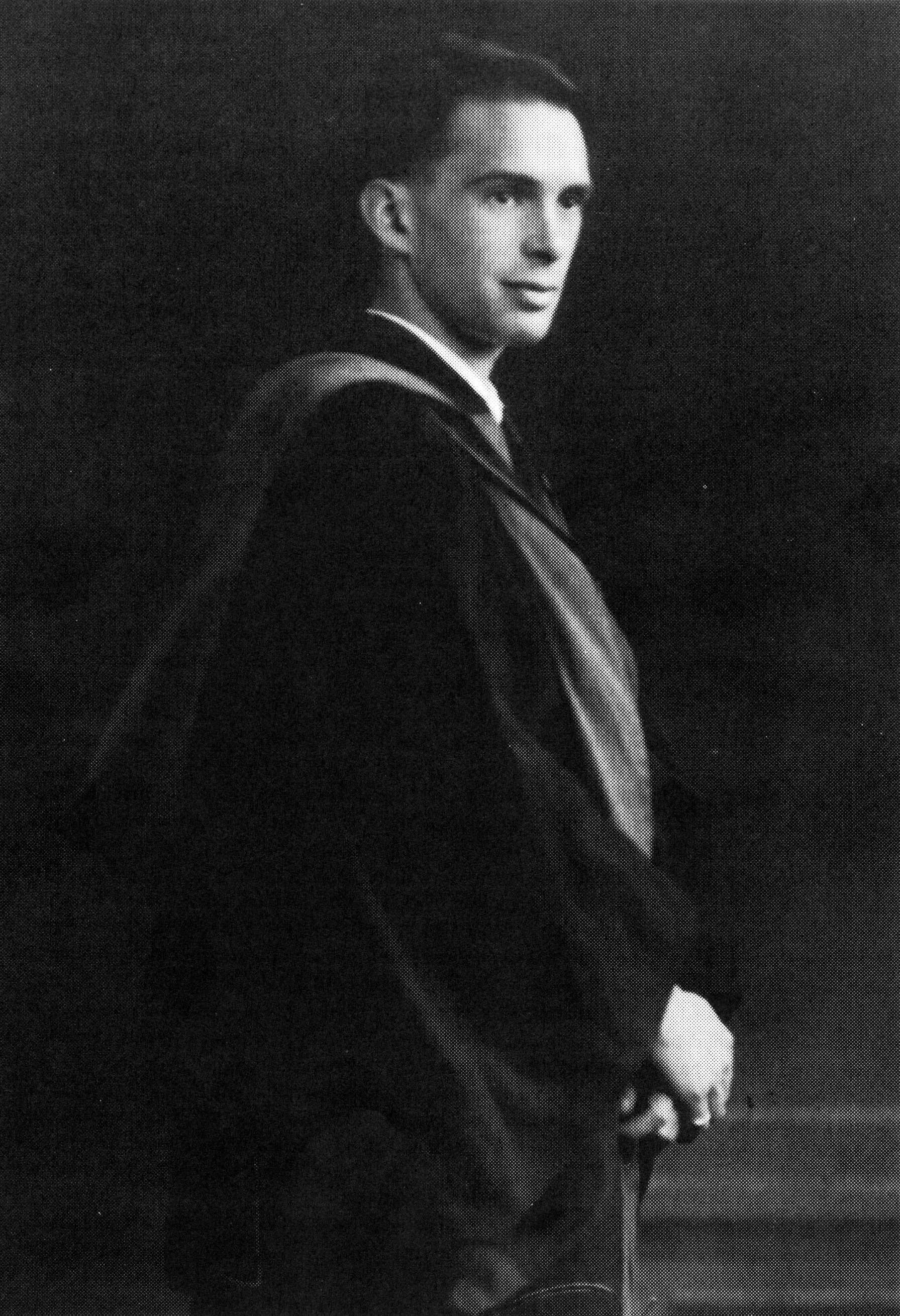}
 \includegraphics[scale=0.29]{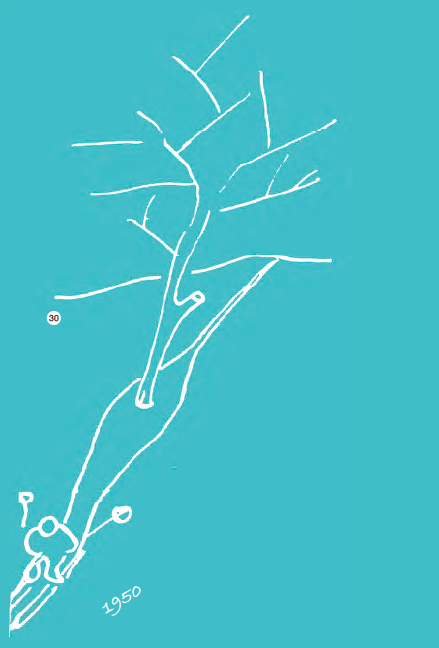}
  \caption{{Left panel:}  Max Born and Werner Heisenberg, in Glasgow, 1947, courtesy of CERN Documentation Service. {Center panel:}  Bruno Touschek's 1949 D.Phil. official photograph  from University of Glasgow, from   the Bruno Touschek papers at  \RSUPD. {Right panel:}  a drawing by Bruno Touschek in a 1952 letter to his father from Glasgow, \copyright Touschek Family,   rendering  by A. Ianiro, no reproduction allowed.}
  \label{fig:toga-BornHeis-sciatore}
 \end{figure}
  Bruno's formation as a theoretical physicist was completed in Scotland, where he earned  his D.Phil. in 1949  from the University of Glasgow, and became a collaborator of   another  theorist who had shaped pre-war nuclear science, Max Born,  then Tait Professor in the University of Edinburgh, {seen in the left panel in Fig.~\ref{fig:toga-BornHeis-sciatore}}.\footnote{Max Born was the 1954 Nobel Prize recipient ``for his fundamental research in quantum mechanics, especially for his statistical interpretation of the wavefunction", \url{https://www.nobelprize.org/prizes/physics/1954/born/facts/}.} For Born,   Bruno curated the  proofs of his book {\it Atomic Physics} and prepared an appendix to it \cite[p.218-222]{Pancheri:2022nzs}. 
   Shortly after his doctorate, Fig.~\ref{fig:toga-BornHeis-sciatore},  Bruno also wrote a remarkable paper with Walter Thirring \citep{Thirring:1951cz} 
about the infrared catastrophe and the covariant formulation of the Bloch and Nordsieck theorem \citep{Bloch:1937pw}, a life-long interest of Touschek's \citep{Etim:1967mxo}. { He remained  in Glasgow until December 1952, and joined the Physics Institute of the University of Rome in January 1953.}
 
 I have gone through the list of  some  great scientists which contributed to Touschek's formation as a theoretical physicist, to contrast the view, commonly held  after his death,  that his  proposal  to build a new type of  accelerator had    come almost casually after a seminar by Wolfgang Panofsky, the director of the Stanford Linear Accelerator Center. While  this was  a  moment in time when things were set  in motion as detailed in \citep{Bonolis:2023lmh},   it was Bruno's theoretical physics knowledge that gave him the vision and determination to pursue the  idea  of what could and should be done to further probe nature's deeper world. Of course, as often repeated, he also had  first-hand knowledge of the mathematics involved in building an electron accelerator 
 \citep{Amaldi:1981be}
 and when he proposed to build AdA, he needed to convince  his fellow scientists in the Frascati National Laboratories that it would be worth doing, investing money and people, engineers and physicists, the director of the Laboratory and the President of  the National  Institute for Nuclear Physics (INFN),  who had to approve the budget and present the request to the Ministry of Industry for providing the funds. His success in this first phase between February and March 1960 came from his deep conviction that the machine he proposed to build was the  future of particle physics.  This same deep faith  carried him and his team through the three years that it took to prove the feasibility of electron positron collisions.

 But of course, theoretical physics  is not sufficient. As the  case of electron positron physics shows, it is not enough to know the physics potential, which Touschek did, as one must also know that it can be done,  which he did as well, having learnt the ways electrons in a magnetic field can be controlled   when  he  was   in Germany,  during the war, earning  his living by assisting the Norwegian scientist \RW \ to build a 15 MeV betatron for military uses \cite[Chap. 5]{Pancheri:2022nzs}. And yet, to reach deeper into the structure of matter, as electron-positron colliders aimed to do, a single European country on its own, even if equipped with a national laboratory as Frascati was, could not overcome the limitations  of space, financial investments, human resources and up-to-date  transfer of knowledge. How post-war Europe provided all this and gave birth to electron positron colliders will be highlighted in the next section. 
       
\section{The making of AdA and the European dimension}
\label{sec:ada-Europe}
  AdA was  observing
 electrons or positrons circulate  in the same magnetic ring in February 1961 \citep{Bernardini:1962zza}, but it took three more years before the definite  proof of the   feasibility of electron positron colliders  as an effective tool for  high energy physics experiments.

The result was obtained through both theoretical work in the University of Rome \citep{Cabibbo:1960zza, Altarelli:1964aa}  
and the  joint experimental effort \citep{Haissinski:1965,Haissinski:2023ene}  between the Frascati National Laboratories and the
 \LAL \  in Orsay, where 
 AdA was taken to improve its performance, thanks to 
 the higher photon yield of the LINAC \citep{Pancheri:2018xdl}. {It is part of AdA's legend that}  since strict custom regulations  between Italy and France did not allow for free movement of goods or people, it took the intervention of  
 Francis Perrin, {\it Haut Commissaire \` a l’
Energie Atomique au CEA}, one of the founders of CERN  who had worked with Fr\'ed\'eric Joliot
on nuclear chain reactions and was very  influential, before the truck carrying AdA  and  all its working vacuum pumps,   could enter France. As Carlo Bernardini recalls,  in July 1962, a direct telephone consultation between ministers in France and Italy, became necessary in order for the newly constructed storage ring AdA, to cross the existing custom barriers \citep{Bonolis:2018gpn}. \footnote{The episode is also recalled by the French physicist Fran\c ois Lacoste, in a 2013 interview, at Orsay, for the docu-film {\it Touschek with AdA in Orsay.}} 
 
 {The transfer had been envisaged during Marin and Charpak's  visit to Frascati in July 1961, and was discussed in encounters and letters exchanged by  the two laboratories,  between September 1961 and July 1962.}\footnote{{In 2011, copies of letters and other supporting material describing these  exchanges  were  sent  by e-mail  to this author  by the late \JH, who had digitalized them  from the Archives in Orsay \LAL. Details of their content appeared in  \citep{Bonolis:2011wa},  \citep{Bonolis:2018gpn} and \cite{Pancheri:2022nzs}.}}  {This  international transfer of people and machinery  
   had its origin through  the special role of CERN for Europe's scientific exchanges.} 

The starting point {for the transfer had taken}  place in Geneva,   at  the CERN  1961  Conference on Theoretical aspects in High Energy Physics, during the session on  Electromagnetic Processes \citep[p. 57]{Bell:1961gi}, where Touschek presented the Frascati electron-positron storage ring under construction, AdA,  and the ADONE project for a much more powerful collider, {proposed in \citep{Amman:1961}, which he had dreamed}  
 could  reach  an unthinkable energy,   $E_{cm}=3.0$ GeV.\footnote{{Touschek proposed to build ADONE in a memo dated November 9th, 1960, 
  \citep[p.397]{Pancheri:2022nzs}. In February the proposal was formally prepared as LNF note \citep{Amman:1961}. }} ADONE  became reality  operating with two beams in 1969, right away  observing an anomalous production of hadronic particles \citep{Lubkin:1973}.

 Following  and complementing  Touschek's talk {at the CERN 1961 Conference}, was his colleague Raoul Gatto, \citep{Bonolis:2023lmh}, another first class theorist from the University of Rome, who gave a talk about the physics prospects at such machine. In the CERN audience, the idea sparked interest while news of Frascati's bold step forward reached both the USSR and France.  

\begin{figure}
 \centering
  \includegraphics[scale=0.146]{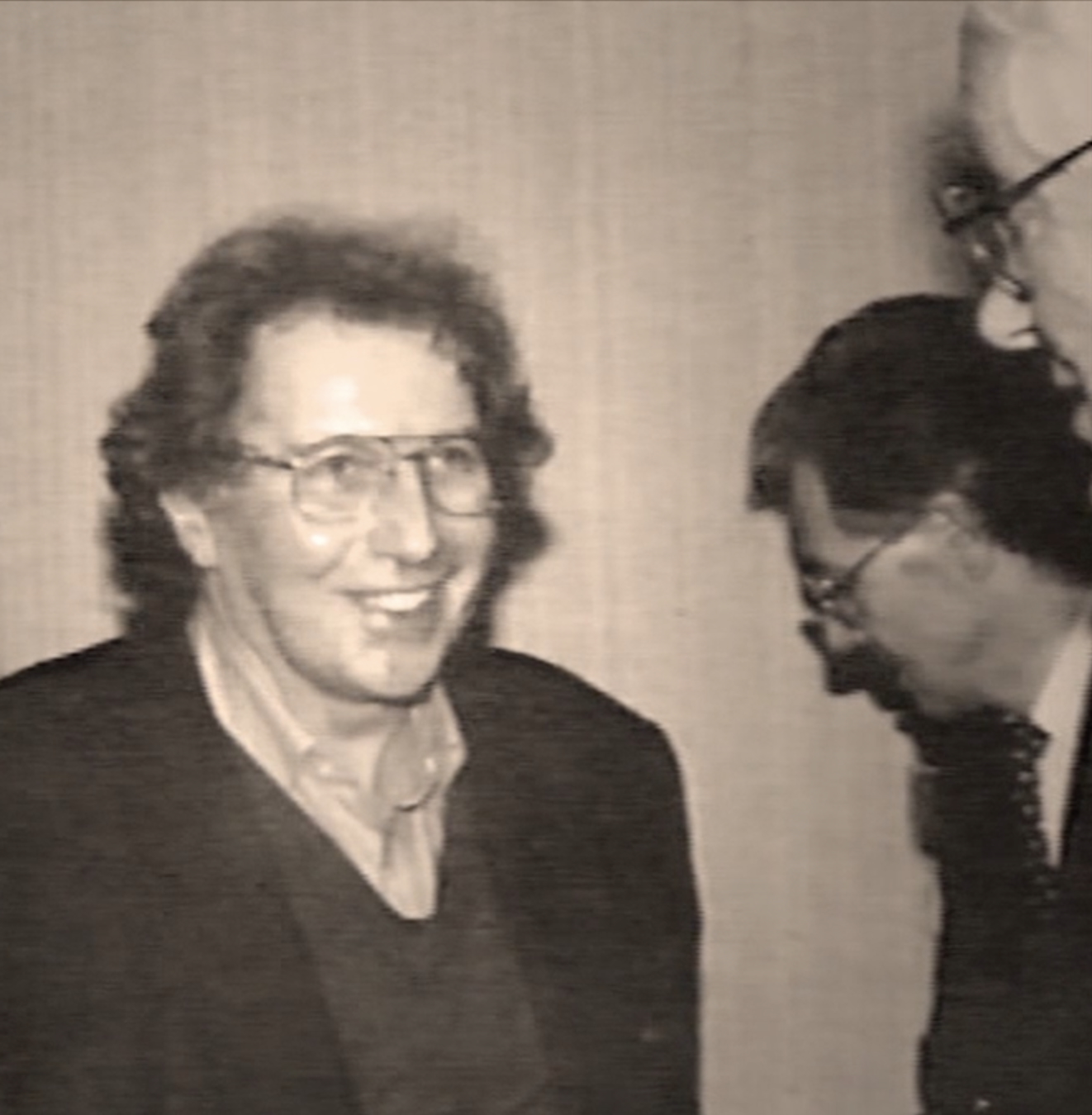}
   \hspace{0.1cm}
 \includegraphics[scale=0.11]{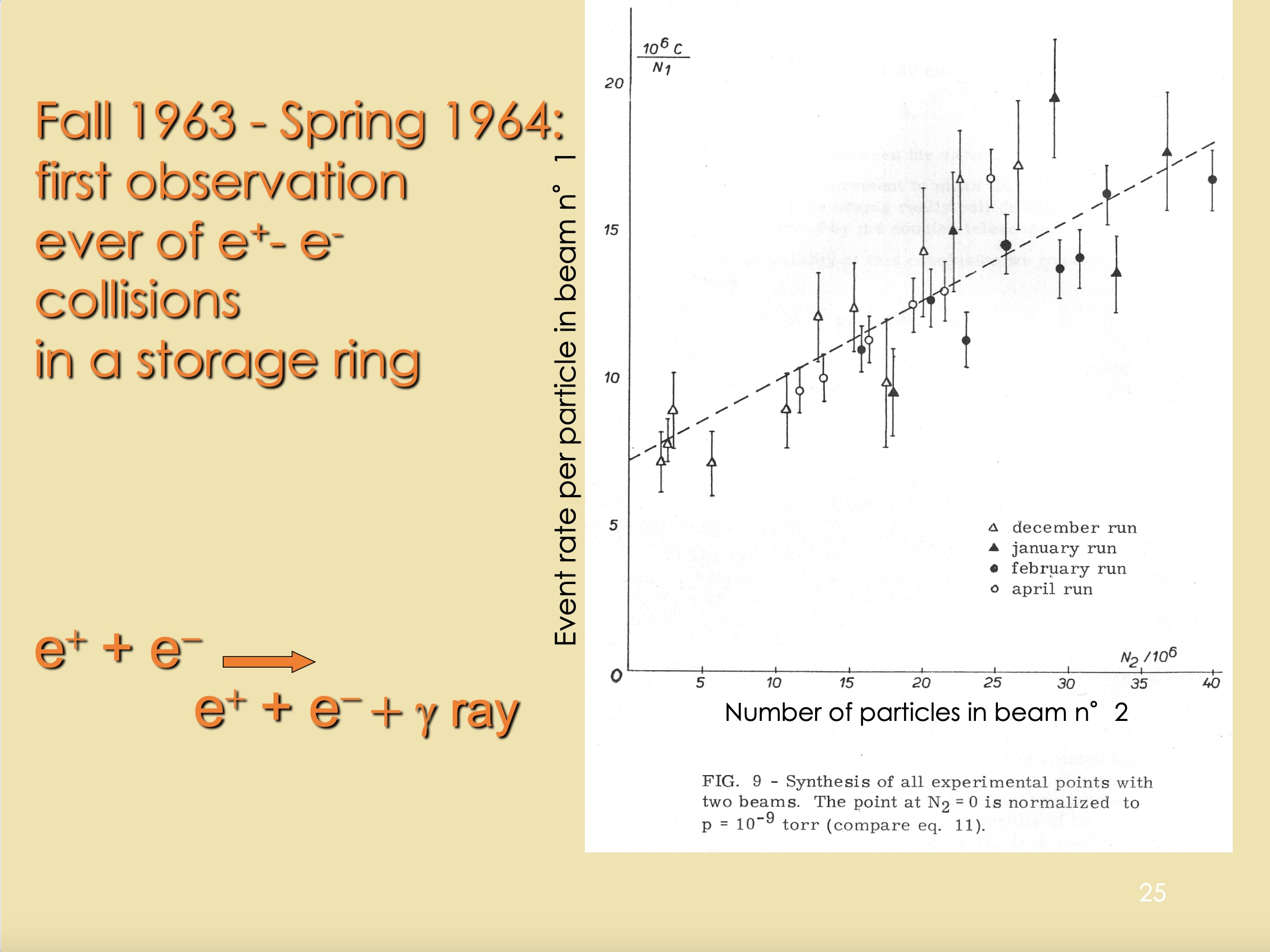}
 \hspace{0.1cm}
\includegraphics[scale=0.162] {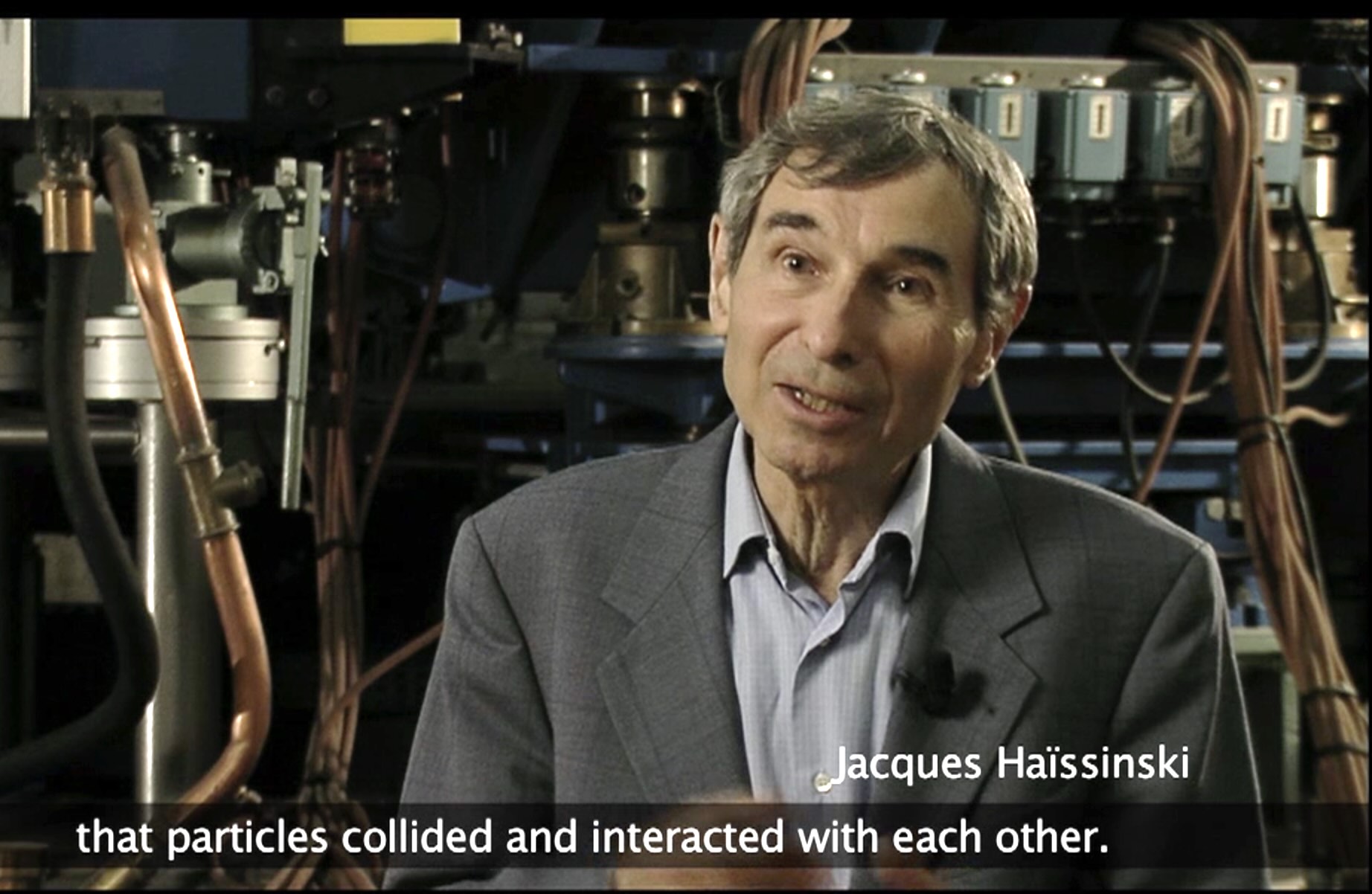}
 \caption{{Left panel:} \PM \ with Georges Charpak (barely seen at right in the first panel)  in the late '70s, photo courtesy of \JH. {Center panel:} transparency presented  by  \JH \ in 2020 in Rome \citep{Bonolis:2023dle}, 
 with   the plot  proving that AdA had observed collisions during four  runs in the \LAL \ d'Orsay, now IJC-Lab, from Carlo Bernardini's Papers in \RSUPD. {Right panel:} still shot of \JH \ in the docu-film {\it Touschek with AdA in Orsay}, \copyright E. Agapito and INFN.}
 \label{fig:MarinCharpak-collision-JH}
 \end{figure}

 In France,   Pierre Marin,  Fig.~\ref{fig:MarinCharpak-collision-JH},  from  the Orsay  \LAL  \  was looking for new research directions and went to Frascati to see what was happening \citep[p. 46]{Marin:2009}. This visit started the international collaboration which
  one  year later, brought  AdA to be transported to Orsay, where collisions were observed within less than  two years. The laboratory work and the details of  this ground-breaking observation were described in his  {\it Th\`ese d'\'Etat} by  \JH, Fig.~\ref{fig:MarinCharpak-collision-JH}, later to be one of the protagonists of the construction of the French collider ACO, {\it Anneau de Collisions d'Orsay}, which started operations in 1968 \citep{Augustin:1968zz}.  
 In Novosibirsk,   the  theorist V.N. Baier, in late October 1959,  had   convinced the laboratory director, G. Budker, of the {advantage} in building an electron positron collider \citep{Baier:2006ye}, but progress had been slow because of  inherent difficulties and the isolation of the Russian scientists. 
 The news from Frascati encouraged them,   and their  electron-positron collider, VEPP-II, could be seen to be in advanced state of preparation in 1963 \citep{Marin:2009}, 
 \citep[p.381-383]{Pancheri:2022nzs}. 
By this time, new collider projects had been approved, in Italy with ADONE, in France with ACO,  at CERN with the ISR, Intersecting Storage Rings, while in the USA the electron-electron project was already in construction. 

The impact of Touschek and Gatto's  talks at CERN in June 1961 is not limited to the interest it generated   in Orsay.  
    The Italian-French collaboration, which  demonstrated the feasibility of electron positron colliders, was started by a visit to Frascati, one month after the June 1961 meeting at CERN \citep{Bonolis:2018gpn},
 but  Pierre Marin  did not go  alone. He went there, he writes,  with  Georges Charpak, Fig.~\ref{fig:MarinCharpak-collision-JH}, who had joined CERN in 1959, where discussions about future programs had started since the PS had began operations. Storage rings were part of on-going discussions and the two presentations by the Italian scientists,
 followed by Charpak's visit to Frascati, signalled the beginning of a new era. 
 
 CERN had experience with accelerating proton beams. It is therefore no accident that while Frascati and the \LAL \ pursued a collaboration about electron positron colliders, at CERN the interest in storage rings veered to  $pp$ collisions \citep{Hereward:1960zz}, and the ideas of what would become the Intersecting Storage Rings  would be  studied by  a group of ``enthusiasts" \citep{Hereward:1961wxa}    led by  Kjell Johnsen, who invited  Touschek to join them \citep{Bonolis:2018gpn}.\footnote{K. Johnson solicited  Touschek to  join the next meeting of the group through Fernando Amman, who was in charge of the nascent ADONE project,   as we learn from Touschek's  letter to Johnsen on February 21st, 1962, Touschek Papers in \RSUPD.} Touschek
 accepted the invitation but he was very busy with planning for AdA's transfer to Orsay, and there is no information about his going to CERN in Spring 1962. 
 
{AdA arrived in Orsay   in July 1962, to be  installed in Hall-500 of the  \LAL\ in November. Less than  two years later,  proofs of feasibility of Touschek's electron-positron experiment were submitted for publication \citep{Bernardini:1964lqa}. 
 By this time, news about Frascati's results had led to proposal and approval of   projects for electron-positron colliders such as the Anneau de Collisions d'Orsay (ACO) in France, ADONE in Italy, VEPP-2 in the USSR.}  In the 1970s, the traditional way of accelerating particles to hit a fixed target was slowly abandoned in favor of collisions between moving beams and, in the second part of last century,  different  types of particle colliders were designed and built. New facilities hosted  new  accelerators \citep{Amaldi:2015iia}
  for   electrons clashing against positrons, 
  protons against other protons or against anti-protons, 
  positrons against protons,   heavy ions against protons, 
  or other heavy ions. Reaching higher and higher energies  allowed the experimental discovery of  particles which had been 
  proposed by the theoretical physics community to be the building bricks of the Standard model of elementary particles \citep{Treille:2025myk}. 
  
   Thus, in a joint march toward the unknown,  {sponsored and encouraged by international and intergovernmental exchanges}, 
    theorist and experimentalists led and followed the development of ever more powerful particle accelerators, until 
   the {LHC, in 2012},  closed the first chapter of the Standard Model of Elementary Particles, with the discovery of the Higgs boson, born as a 
 theoretical physics mechanism proposed in {1964} and searched for, in its particle form,  through   many decades. Such is the power of theoretical physics.   

\section{{Conclusion}}
{The story I outline in this article captures the making of a revolution in particle accelerator physics in the second half of last century. {Such a  revolution  made  particle colliders the major discovery tool in high energy physics, and} was made possible by international collaboration, interregional and intergovernmental treaties and  exchanges.}

{ The vision which guided 
scientists  in 1950 led the European governments to establish an organization such as CERN, the Conseil \'Europe\`en pour la Rech\`erche Nucl\'eaire which would reunite and guide European scientists towards post-war reconstruction.
 The  substantial costs carried  by   competing scientifically with the US was provided jointly by the Member States,  in proportional to  national budgets. Italy and France were founding   members, but on opposite sides, as France was among the WWII victors, and  Italy  on  the losing side.}

{CERN's scientific project was the  study of nuclear physics, to probe interactions among matter constituents. Around such a vision, European  scientists built a powerful proton accelerator, and, in parallel,   two national laboratories and two modern type accelerators. For Italy, which the war had tragically divided into two parts in 1943,  the Frascati synchrotron became a symbol of unity
and exceptional progress, led by Edoardo Amaldi, one of the founders of CERN, and President of a newly born National Institute for Nuclear Physics in 1952. The next step,  in  developing the extraordinary discovery tool of  particle colliders, came from the vision of theoretical physicists. Central to the new step was  the CPT theorem and 
Touschek's unshakeable faith in its validity which led to the birth of  electron positron physics. A new field of physics was thus born through the 1960s 
in Europe, the USSR  and  the USA, where  both   electron-electron and electron-positron colliders were built \citep{f4b42b80-626c-3590-ac25-13f5decafdf3}.}
{The scientific revolution brought by elementary particle accelerators is now returning its due to society, as  seen in the pervasive usage of particle beams for medical and industrial purposes, which are presently routinely performing in most advanced countries with  beneficial political and societal effects.
 Such benefits must be extended to other, less developed,  regions and the story told in this article could provide inspiration for  
the scientific community in the greater Caribbean region to carry their voice to the governments for the creation of a regional alliance towards a large project such as the Greater Caribbean Synchrotron, in a guiding vision to include both the scientists and the political world and their interaction.}

{Scientists, and, in particular  theoretical scientists,  are used to work on a mental scale beyond their life-time, thus their vision may overtake  present tragedies and conflicts, and may be the ideal conveyors for carrying  their projects to government attention.   If  the political world, with its different vision can rise to the challenge,  great steps  to society's benefit can occur.  }

{While it is true that history does not repeat itself, it can teach and encourage to dream and to plan and  make things, otherwise and apparently impossible, to become reality, as in cases such as the one recollected here. If  this article  can provide inspiration  to pursue  the construction  of the Greater Caribbean Light Source Project, the story of Touschek's small machine and the Frascati synchrotron will return to life  and be of use to future generations.  }
 
  \section*{Acknowledgments}
{ I acknowledge many useful suggestion from the Editorial Office in the preparation of this article. } Thanks are due to  the Frascati Laboratory Documentation Service and the library staff for their support and advice.  I  am grateful to Yogendra Srivastava for a careful reading of this manuscript  and to Galileo Violini for arising my interest in  the role of intergovernmental cooperation in the progress of modern science and  am indebted to Luisa Bonolis for helpful  discussions  and sharing of documents cited in this note.  I thank my French colleagues from the IJC-Laboratory  in Orsay for the invitation to the 2025  {\it Colloque Ha\"issinski}, which inspired this note and where I presented some of its content, in memory  of  \JH.\footnote{\tr{See \url{https://www.sfphysique.fr/evenement/colloque-jacques-haissinski/}.}}

\bibliographystyle{apalike-refs}
\bibliography{BT-power-2}
\end{document}